\documentclass[conference]{IEEEtran}
\IEEEoverridecommandlockouts
\usepackage{cite}
\usepackage{amsmath,amssymb,amsfonts}
\usepackage{algorithmic}
\usepackage{algorithm}
\usepackage{graphicx}
\usepackage{textcomp}
\usepackage{xcolor}
\usepackage{float}

\def\BibTeX{{\rm B\kern-.05em{\sc i\kern-.025em b}\kern-.08em
    T\kern-.1667em\lower.7ex\hbox{E}\kern-.125emX}}

\begin{document}

\title{Enhanced Sliding Window Superposition Coding for Industrial Automation
\thanks{The paper is sponsored in part by the National Key R\&D Program of China No. 2020YFB1806605, by the National Natural Science Foundation of China under Grants 62022049 and 62111530197, and by Hitachi.}
}

\author{\IEEEauthorblockN{Bohang Zhang, Zhaojun Nan, Sheng Zhou, Zhisheng Niu}
\IEEEauthorblockA{Department of Electronic Engineering \\Beijing National Research Center for
Information Science and Technology,\\
Tsinghua University, Beijing 100084, China\\
Emails: \{zhangbh20@mails., nzj660624@mail.,
sheng.zhou@, niuzhs@\}tsinghua.edu.cn.}
}

\maketitle

\begin{abstract}
The introduction of 5G has changed the wireless communication industry. Whereas previous generations of cellular technology are mainly based on communication for people, the wireless industry is discovering that 5G may be an era of communications that is mainly focused on machine-to-machine communication. The application of Ultra Reliable Low Latency Communication in factory automation is an area of great interest as it unlocks potential applications that traditional wired communications did not allow. In particular, the decrease in the inter-device distance has led to the discussion of coding schemes for these interference-filled channels. To meet the latency and accuracy requirements of URLLC, Non-orthogonal multiple access has been proposed but it comes with associated challenges. In order to combat the issue of interference, an enhanced version of Sliding window superposition coding has been proposed as a method of coding that yields performance gains in scenarios with high interference. This paper examines the abilities of this coding scheme in a broadcast network in 5G to evaluate its robustness in situations where interference is treated as noise in a factory automation setting. This work shows improvements of enhanced sliding window superposition coding over benchmark protocols in the high-reliability requirement regions of block error rates $\approx 10^{-6}$. 
\end{abstract}

\section{Introduction}
With the release of 5G and the subsequent rollout of this network, the wireless industry is beginning to realize that 5G may not be as much an era for interpersonal communication but rather for communication between machines. The applications of Ultra-Reliable Low Latency Communications (URLLC) are of vital importance as it aims to create a balance between both reliability and latency. The requirements of URLLC is usually $10^{-4} \text{to} 10^{-6}$ Block Error Rate (BLER) at 1-10ms latency. Traditional methods of improving communications usually come at the cost of reliability or latency, i.e HARQ can improve upon the reliability of the communications link at the cost of latency \cite{HARQ} while simply shortening the error correcting code will decrease latency at the cost of error rate \cite{Poor2018}. 
With this proliferation of URLLC applications, the factory automation scenario appears to be an area of great interest as it combines URLLC with the rise of machine-based communication. In particular, coding schemes are important for the attainment of URLLC as coding can determine the fundamental limits of the system. Research by others has identified LDPC as a good benchmark when dealing with the factory automation setting \cite{LPDCFA}. 

Most of the previous works are mainly dealing with channels where individuals have considered a system under Orthogonal Multiple Access (OMA). This may not be a suitable scheme for URLLC as one major source of delay in communications systems is the multiple access scheduling. As such, work by others has considered the applications of Non-Orthogonal Multiple Access (NOMA) for URLLC \cite{NOMAURLLC} \cite{anlzGFA}. This application of NOMA also extends into the factory automation setting as well\cite{NOMAFA}. A major challenge of NOMA is the coding scheme used to deal with the interference from other uses and work having been done on evaluating associated coding schemes \cite{GFASIC}. On the issue of interference, Sliding Window Superposition Coding (SWSC) had been proposed by others for use inside the interference channel\cite{Wang2020} \cite{SWSCOG}. Whereas previous works of SWSC are mainly focused on the achievable rate at a Block Error Rate of 0.1, this work will focus on regions of high-reliability requirement where message error rates are $\approx 10^{-4}$ to $10^{-6}$. Another challenge of URLLC and using LDPC is the issue of error floor \cite{LDPCErrorFloor} at short block length \cite{LDPCBlockLength}. This work will take this into account when constructing associated simulations to offer a complete comparison.

The main contributions of this work are as the following:
\begin{enumerate}
\item Propose a new decoding method for SWSC called Enhanced SWSC (ESWSC) that is able to address the issue of error propagation within SWSC to increase performance. This is done while maintaining the same coding efficiency as SWSC. 

\item Evaluate the performance of SWSC and ESWSC through simulations and observe the impacts of block length, number of blocks, and block ratio $\alpha$ (for ESWSC) on the overall performance of the codes.

\end{enumerate}

This work will first propose an improved decoding method for SWSC called Enhanced SWSC (ESWSC) in Section II. The benchmark protocols will be LDPC and a modified version of LDPC (mLDPC). These protocols will also be outlined in Section II. The simulation parameters are discussed in Section III where the results are also analyzed. The conclusion in Section IV will summarize the results obtained in the previous sections. 

\section{System Model}
An indoor factory automation situation is used where the information for $N$ receivers is sent in a broadcast manner for downlink. The downlink message is sent through the standard 5G specifications on the data plane \cite{release15}. The broadcasted message contains the information intended for all receivers and each round of broadcast is considered successful if all of the $N$ receivers are able to decode their respective messages. The proposed algorithm of ESWSC will be compared against benchmark SWSC and versions of LDPC. 

\subsection{SWSC}
\begin{figure}[H]
\centerline{\includegraphics[scale=0.25]{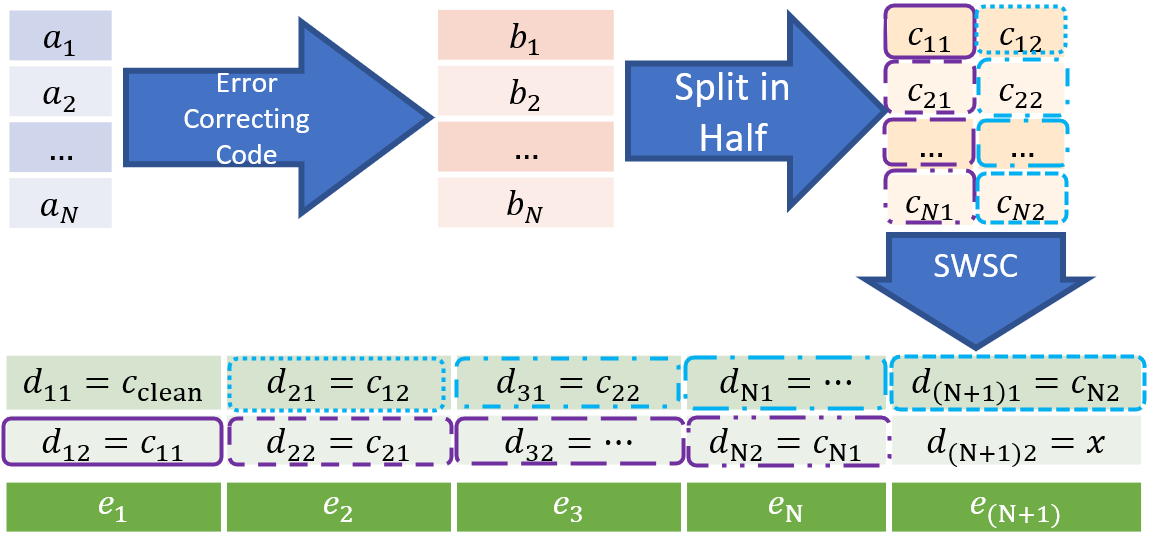}}
\caption{SWSC Encoding}
\label{fig: SWSCEnc}
\end{figure}

 For $N$ total information packets that need to be sent, $a_i$ is an individual information packet while $a=(a_1, ..., a_N)$ denotes all the information bits that are sent. $b_i$ is the coded bits after $a_i$ has gone through error correcting code encoding with $b=(b_1, ..., b_N)$ denotes all the bits after error correcting code.  $c_{i1}$ and $c_{i2}$ are the first and second half of bits in $b_i$ respectively so  $b_i = c_{i1} \mathbin\Vert c_{i2}$ with $c_1=(c_{11}, ..., c_{N1})$ and $c_2=(c_{12}, ..., c_{N2})$ denoting the first and second halves of $b$ respectively. For simplicity's sake, the example in Fig. \ref{fig: SWSCEnc} shows SWSC of 2 layers with $d_{i1}$ and $d_{i2}$ denoting the location within the layer of superposition coding but this can work for any number of layers. The bits in $d_{11}$ will be a known sequence called $c_{\text{clean}}$. For each superposition location pair of $d_{i1}$ and $d_{i2}$, the layers can be modulated together to form $e_i$ with $e=[e_1, ..., e_N]$ is the message that is sent through the channel.

\begin{algorithm}[htbp]
  \caption{SWSC Encoding Scheme}
  \label{alg:SWSC Encoding Scheme}
  \begin{algorithmic}[1]
     \STATE Assume $N$ original information packets are to be sent with each packet being $a_i$ .
     \STATE $a_i$ becomes $b_i$ from error correcting code.
     \STATE Split $b_i$ into $c_{i1}$ and $c_{i2}$
     \STATE 
     \FOR {$i \in 1\rightarrow N $}
     \IF {$i= 1$}
     \STATE   Place a known sequence $c_{\text{clean}}$ in $d1_{1}$ location
     \STATE   Place  $c_{11}$ and $c_{12}$ into  $d_{21}$ and $d_{12}$ respectively
     \ELSE
     \STATE    Place  $c_{i2}$ and $c_{i2}$ into  $d_{i2}$ and $d_{(i+1)1}$ respectively

     \ENDIF
     \STATE bits in $d_{i1}$ and $d_{i2}$ are modulated to form $e_i$.
     \ENDFOR
  \end{algorithmic}
\end{algorithm}

\begin{figure}[htbp]
\centerline{\includegraphics[scale=0.25]{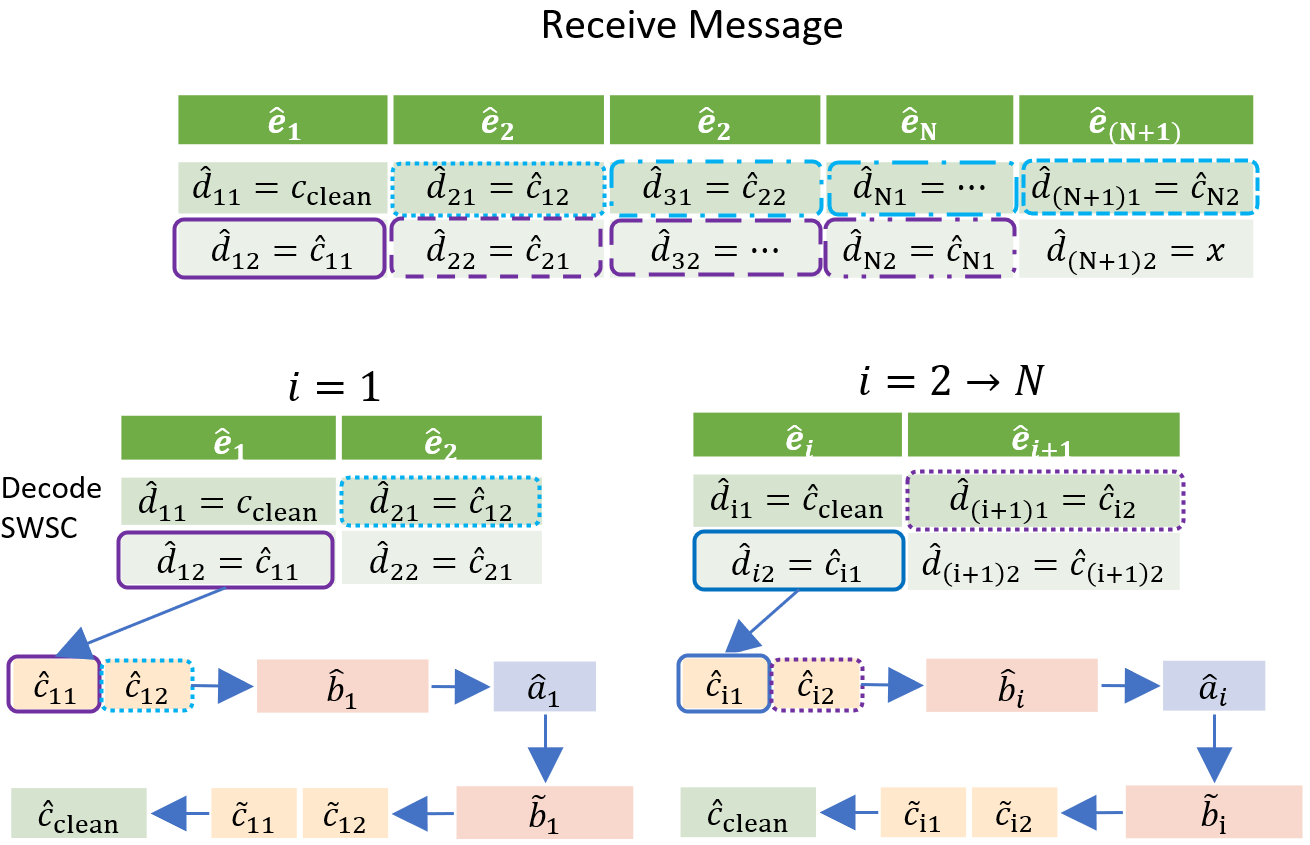}}
\caption{SWSC Decoding}
\label{fig: SWSCDec}
\end{figure}

The decoding of SWSC is the reverse of all the steps in the encoding process. To start, $\hat{e_1}$ will decode into $\hat{d_{11}}$ and $\hat{d_{12}}$ where $\hat{e_i}$ is the received message after $e_i$ goes through the wireless channel. $\hat{c_{11}}$ will have a higher chance of being right due to the $c_{clean}$ in $\hat{d_{11}}$. The existence of a known message will make decoding much easier to complete. Next, $\hat{e_2}$ will decode into $\hat{c_{12}}$ and $\hat{c_{21}}$ respectively. $\hat{c_{11}}$ and $\hat{c_{12}}$ will combine to form $\hat{b_{1}}$ which can obtain $\hat{a_{1}}$. $\hat{a_{1}}$ can form $\tilde{b}_1$ and therefore $\tilde{c_{12}}$ which can serve as the new $\hat{c}_{\text{clean}}$ sequence that is used for further sliding window decoding. This sequence is repeated in the manner outlined in Algorithm \ref{alg:SWSC Decoding Scheme} and shown in Fig. \ref{fig: SWSCDec}. 

\begin{algorithm}[htbp]
  \caption{SWSC Decoding Scheme}
  \label{alg:SWSC Decoding Scheme}
  \begin{algorithmic}[1]
     \STATE $\hat{e_i}$ is the received message after $e_i$ goes through channel
     \FOR   {$i \in 1\rightarrow N $}
     \IF {$i= 1$}
     \STATE $\hat{e_1}$ gets back bits in $\hat{d_{i1}}$ and $\hat{d_{i2}}$ through demodulation
     \STATE The bits in $\hat{d}_{11}$ is a known sequence $c_{\text{clean}}$. This will make the bits in $\hat{d}_{12}$ easier to decode.
     \STATE Demodulate $\hat{e}_{2}$ to get the bits in $\hat{d}_{21}$ and $\hat{d}_{22}$
     \STATE Use $\hat{c}_{11}$ from $\hat{d}_{12}$ and  $\hat{c}_{12}$ from $\hat{d}_{21}$ to construct $\hat{b}_1$
     \STATE Run $\hat{b}_1$ through error correcting code to obtain information packet $\hat{a}_1$
     \STATE $\hat{a}_1$ becomes $\tilde{b}_1$ through error correcting code
     \STATE $\tilde{b}_1$ splits into $\tilde{c}_{11}$ and $\tilde{c}_{12}$
     \STATE $\tilde{c}_{12}$ is placed in $\tilde{d}_{21}$ which serves as the new clean sequence $\hat{c}_{\text{clean}}$
     \ELSE
     \STATE $\hat{e}_i$ gets back bits in $\hat{d}_{i1}$ and $\hat{d}_{i2}$
     \STATE For any $\hat{e}_i$, the sequence in $\hat{d}_{i1}$ location is assumed to be $\hat{d}_{\text{clean}}$ from the previous coding step. This will make the bits in $\hat{d}_{i2}$ easier to decode
     \STATE Demodulate $\hat{e}_{(i+1)}$ to get the bits placed in $\hat{d}_{(i+1)1}$ and $\hat{d}_{(i+1)2}$
     \STATE Use $\hat{c}_{i1}$ from $\hat{d}_{(i)2}$ and  $\hat{c}_{i2}$ from $\hat{d}_{(i+1)1}$ to construct $\hat{b}_i$
     \STATE Run $\hat{b}_i$ through error correcting code to obtain information packet $\hat{a}_i$
     \STATE $\hat{a}_i$ becomes $\tilde{b}_i$ through error correcting code
     \STATE $\tilde{b}_i$ forms $\tilde{c}_{i1}$ and $\tilde{c}_{i2}$
     \STATE $\tilde{c}_{i2}$ placed in  $\tilde{d}_{(i+1)1}$ to be used as new clean sequence $\hat{c}_{\text{clean}}$ for the next sliding window
     \ENDIF
     \ENDFOR
  \end{algorithmic}
\end{algorithm}

The major advantage that SWSC offers over current protocols is the ability to take advantage of the mutual information between the pieces of the code word from sliding window blocks. This \textit{non-i.i.d} nature of adjacent blocks allows for better decoding as the coding chain is constantly able to create a "clean" message that can aid in the decoding of future blocks.

\subsection{ESWSC}
\begin{figure}[htbp]
\centerline{\includegraphics[scale=0.25]{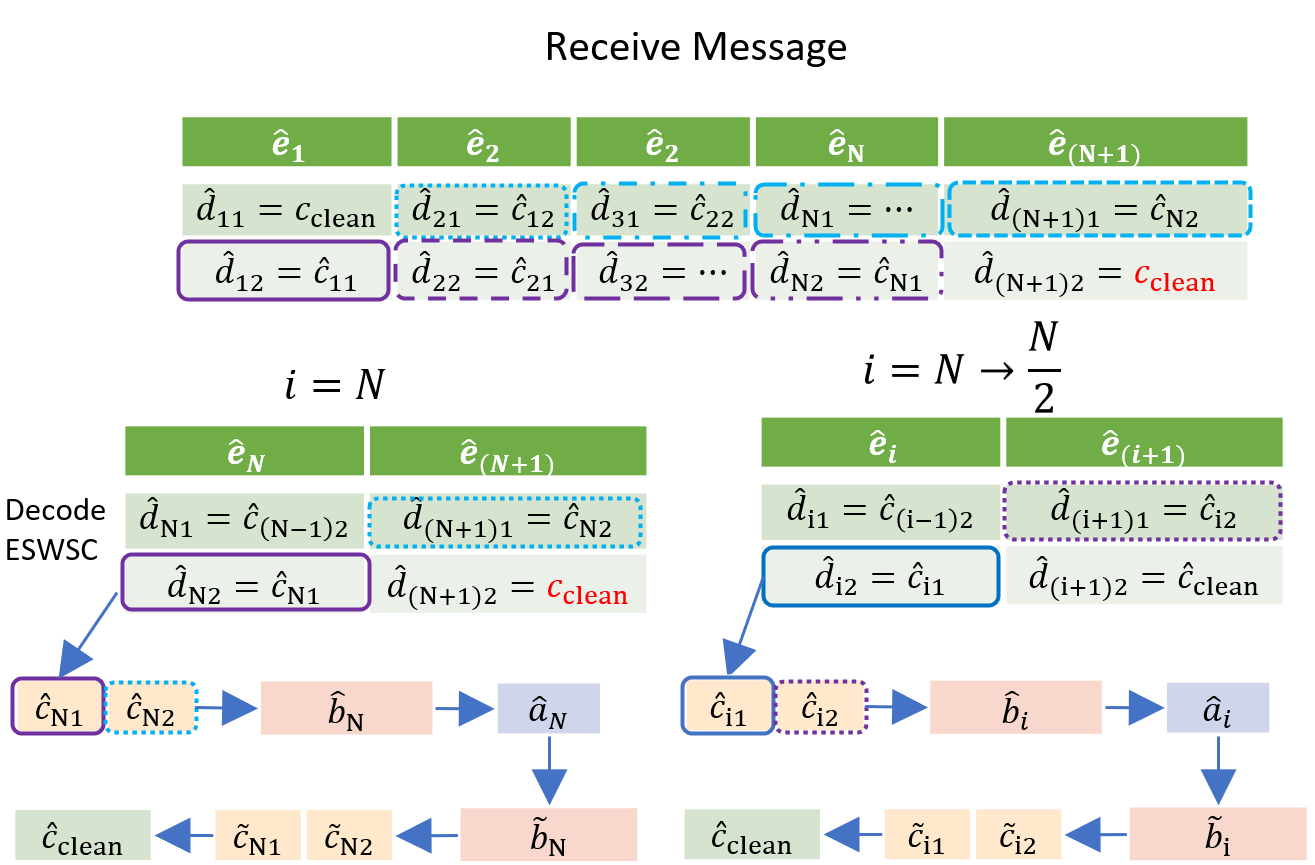}}
\caption{ESWSC Decoding}
\label{fig: ESWSCDec}
\end{figure}

Both the strength and the weakness of SWSC lie in the \textit{non-i.i.d} nature of the adjacent information packets during decoding. The successful decoding of the current message is dependent on the previous message being decoded successfully. This has an issue of error propagation as the unsuccessful decoding of a single package results in the rest of the code being decoded incorrectly. In order to combat this, ESWSC is proposed as an alternative method to decode the SWSC coding chain. While the first portion of the message is decoded in the same manner. A ratio of blocks $0\leq \alpha \leq 0.5$  starting from the back of the SWSC will be decoded in reverse. The $c_{N2}$ sequence of bits will be decoded first aided by the clean message located at $d_{(N+1)2}$ before being combined with $c_{N1}$ to go through error correcting code. This is shown in Fig. \ref{fig: ESWSCDec}. When $\alpha =0$, the code will be decoded in the same manner as SWSC while  $\alpha =0.5$ will be the ideal method to maximize decoding success. The reason for $\alpha$ to have a range of values is that some of the information packets may be latency limited within the frame. This requires a set ordering of the SWSC and $\alpha$ can be adjusted accordingly. In general, most situations are not overall latency constrained but this work takes into account the effect of $\alpha$ in its simulations.

\subsection{Message Error Rate}
The Message Error Rate (MER) can be calculated by finding out $P(a \neq \hat{a})$. In order to evaluate the performance of ESWSC and SWSC under this definition, a number of probabilities will be used. The first probability is the probability of success for the first information packet to be successfully decoded during SWSC or ESWSC decoding $P_\text{clean}=P(a_1=\hat{a_1})$. For any other information packets $a_i$, the probability of successfully decoding it given the previous message is decoded successfully is given by $P_\text{assumed}=P(a_i=\hat{a_i}|a_{(i-1)}=\hat{a}_{(i-1)})$.
As a function of BLER, the MER looks like:
\begin{equation}
\text{MER}=1-{(1-\text{BLER})}^N
\end{equation}

The BLER equivalent for some MER values are shown below in Table \ref{tab MERBER}.

\begin{table}[htbp]
\caption{Approximate BLER values for MER values}
\begin{center}
\begin{tabular}{|c|c|c|c|}
\hline
\textbf{MER}&{\textbf{N=20}}&{\textbf{N=50}}&{\textbf{N=100}} \\
\hline
$10^{-1}$ & $\approx 5.25\times 10^{-3}$ & $\approx 2.10\times 10^{-3}$ & $\approx 1.05\times 10^{-3}$\\
\hline
$10^{-2}$ & $\approx 5.00\times 10^{-4}$ & $\approx 2.00\times 10^{-4}$ & $\approx 1.00\times 10^{-4}$\\
\hline
$10^{-3}$ & $\approx 5.00\times 10^{-5}$ & $\approx 2.00\times 10^{-5}$ & $\approx 1.00\times 10^{-5}$\\
\hline
$10^{-4}$ & $\approx 5.00\times 10^{-6}$ & $\approx 2.00\times 10^{-6}$ & $\approx 1.00\times 10^{-6}$\\
\hline
\end{tabular}
\label{tab MERBER}
\end{center}
\end{table}

For a total number of packets $N$, the probability of all information packets being correct for SWSC is given by:
\begin{equation}
P_\text{SWSC}=P_\text{clean} \times {(P_\text{assumed})}^{(N-1)}
\end{equation}

At the same time, the expression for the performance of ESWSC is given by:
\begin{equation}
\begin{split}
    P_\text{ESWSC}={(P_\text{clean} \times {(P_\text{assumed})}^{(N/2-1)})}^{2}\\
    ={(P_\text{clean})}^2 \times {(P_\text{assumed})}^{(N-2)}
\end{split}
\end{equation}
Since it can be known that 
\begin{equation}
P_\text{clean} \geq P_\text{assumed}
\end{equation}
Therefore
\begin{equation}
P_\text{ESWSC} \geq P_\text{SWSC}
\end{equation}
For the purposes of MER
\begin{equation}
\text{MER}_\text{ESWSC} \leq \text{MER}_\text{SWSC}
\end{equation}

\subsection{LDPC Baselines}
The current standard of 5G dictates the use of LDPC code for the data channel and this shall be used as a baseline comparison \cite{release15}. The \textit{i.i.d} nature of LDPC coding makes the modulation stacking order of the code irrelevant so the LDPC will be constructed in a stacked manner that is similar to SWSC but without the sliding window offset to serve as a comparison. When comparing with ESWSC, extra caution needs to be placed on the fact that $\alpha$ will lead to a difference in total latency constraint. This means that the portion of code within the $\alpha$ constraint can be combined together to form a longer block length LDPC code in order to take better advantage of Shannon theorems. This mLPDC allows for a fair comparison with ESWSC as the latency constraints between the methods will be identical for information packets within the message. This is a modification on the works of others \cite{mLDPC} who had demonstrated that concatenation of blocks in the short block length region will yield to improvement in error performance when keeping the overall latency the same.

\section{Simulations}
A simple simulation will be set up where a number of receivers are scattered around a centralized controller. The channel parameters are calculated based on NYUSIM \cite{NYUSim}. The number of required channels was randomly generated using the default parameters of an indoor simulation scenario as shown in Tables \ref{tab NYUChannel} and \ref{tab NYUAntenna}. A total of $N=\text{100}$ blocks are generated and an appropriate number of sensors are selected randomly through a uniform distribution for $N \leq 100$ as required. The exact values of the simulator are shown in Table \ref{tab NYUChannel} and Table \ref{tab NYUAntenna}.

\begin{table}[htbp]
\caption{ NYUSIM Channel Parameters}
\begin{center}
\begin{tabular}{|c|c|}
\hline
\textbf{Value Name}&{\textbf{Value}} \\
\hline
Scenario & InH \\ 
Frequency (GHz) & 28\\
Barometric Pressure (mbar) & 1013.25 \\ 
Humidity (\%) & 50  \\
RF Bandwidth (MHZ) & 800 \\ 
Temperature(\textdegree C) & 20 \\
Polarization & Co-Pol \\ 
Environment & NLOS \\
T-R Separation Distance Lower Bound(m) & 10 \\
T-R Separation Distance Upper Bound (m) & 50 \\
User Terminal Height (m) & 1.5 \\
Base Station Height (m) & 3\\
Number of RX Locations & 100 \\
TX Power (dBm) & 30\\
Spatial Consistency & Off \\
Human Blockage Parameters & Off\\
\hline
\end{tabular}
\label{tab NYUChannel}
\end{center}
\end{table}

\begin{table}[htbp]
\caption{NYUSIM Antenna Properties}
\begin{center}
\begin{tabular}{|c|c|}
\hline
\textbf{Value Name}&{\textbf{Value}} \\
\hline
TX Array Type & ULA\\
RX Array Type & ULA \\
Number of TX Antenna Elements Nt & 4 \\
Number of RX Antenna Elements Nr & 4 \\
TX Antenna Spacing (in $\lambda$) & 0.5 \\
RX Antenna Spacing (in $\lambda$) & 0.5 \\
Number of TX Antenna Elements Per Row Wt & 1\\
Number of RX Antenna Elements Per Row Wt & 1\\
TX Antenna Azimuth HPBW (\textdegree) & 10 \\
RX Antenna Azimuth HPBW (\textdegree) & 10 \\
TX Antenna Elevation HPBW (\textdegree) & 10 \\
RX Antenna Elevation HPBW (\textdegree) & 10 \\
\hline
\end{tabular}
\label{tab NYUAntenna}
\end{center}
\end{table}

For the coding parameters themselves. A standard 5G downlink model will be used with the following values being selected for default coding parameters as shown in Table \ref{tab downlink}. As the downlink 5G data channel requires the usage of LDPC, it will be the coding scheme that is selected and appropriate CRC attachments are all following the associated 5G standard\cite{release15}. In order to ensure the same coding efficiency, the LDPC-based coding schemes will be using a coding rate that is lower than SWSC-based ones that adjusts according to the exact situations to ensure the total number of bits sent for LDPC-based schemes will $\geq$ than those of the SWSC based ones. This is due to the additional half-block length that is needed for SWSC-based schemes. The following default parameters are used except when changes are needed for results.
\begin{table}[htbp]
\caption{5G Downlink Parameters}
\begin{center}
\begin{tabular}{|c|c|}
\hline
\textbf{Value Name}&{\textbf{Value}} \\
\hline
Transport Block Length (bits) & 200\\
Rate of Code (/1024) & 460 \\
SNR (dB) & 16 \\
$\alpha$ & 0.5 \\
$N$ & 100\\
Modulation & QPSK\\
\hline
\end{tabular}
\label{tab downlink}
\end{center}
\end{table}

The optimal way to achieve MIMO (in this case a $4 \times 4$ channel) channel precoding and decoding is through Singular Value Decomposition (SVD) \cite{SVD}. 

The simulations are completed 250 times with the default parameters shown in Table \ref{tab downlink} and the result can be observed below in Fig. \ref{fig: OG}. This can serve as the baseline for comparisons when changing coding parameters. 
\begin{figure}[H]
\centerline{\includegraphics[scale=0.75]{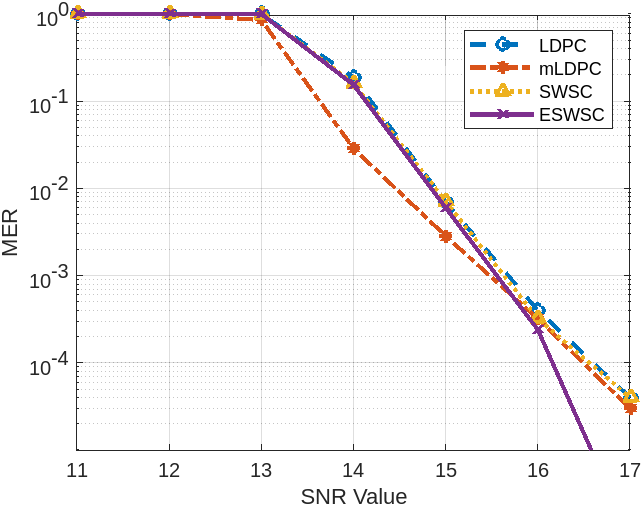}}
\caption{Default Simulation Results}
\label{fig: OG}

\end{figure}

\subsection{The Impact of Code Block Length}
\begin{figure}[H]
\centerline{\includegraphics[scale=0.75]{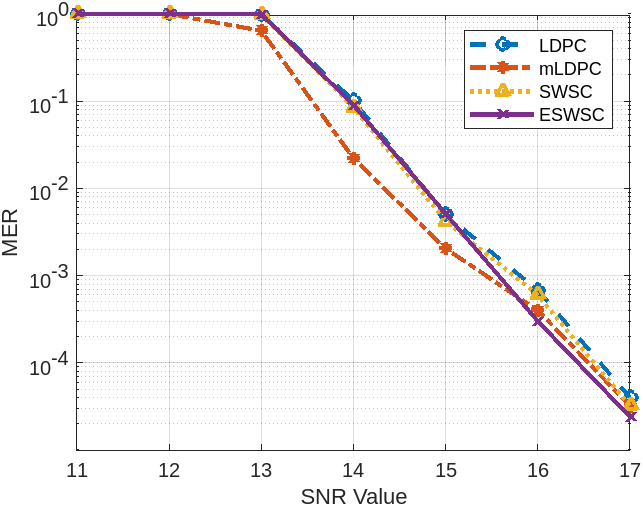}}
\caption{Code Block Length=100}
\label{fig: Code Block}
\end{figure}

When the block length is changed from 200 to 100 bits, the error rate of the various codes changes as seen in Fig. \ref{fig: Code Block}. There is an increase in the success rate of the code at SNR=14 where the requirements for URLLC is not as apparent and where MER is actually lower for shorter block length code. This result is similar to the results obtained by others in which they observed that shorter block length may lead to improvements for LDPC at certain length regions \cite{LDPCBlockLength}. But when attention is focused on the very low MER regions required for URLLC, the impacts of long block length code show how it is able to achieve a lower MER at the same SNR.

\subsection{The Impact of Number of Blocks $N$}
\begin{figure}[H]
\centerline{\includegraphics[scale=0.75]{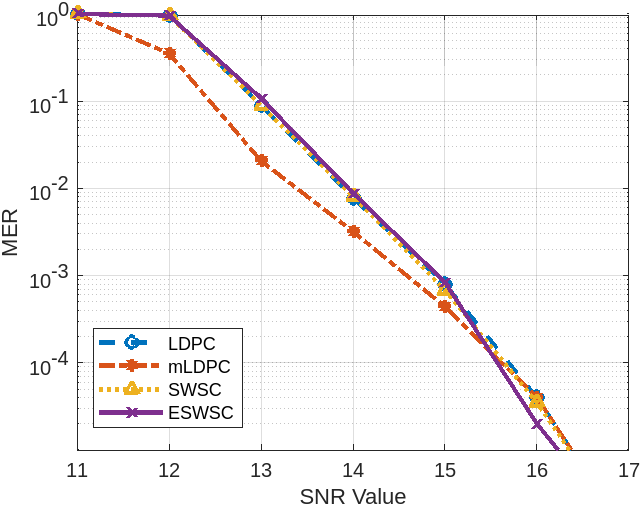}}
\caption{N=20}
\label{fig: N=20}
\end{figure}

\begin{figure}[H]
\centerline{\includegraphics[scale=0.75]{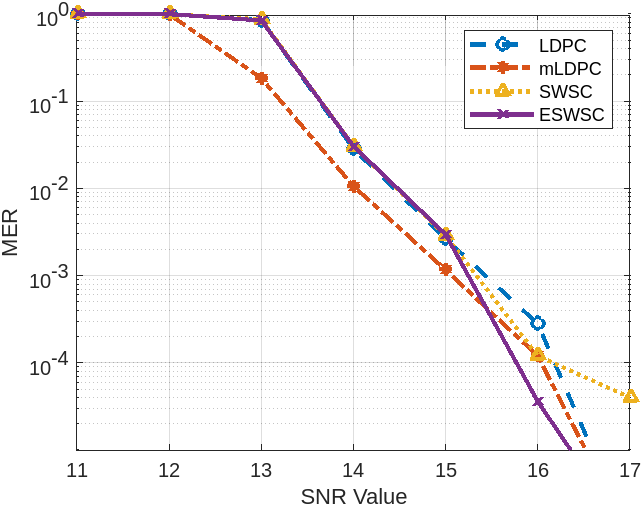}}
\caption{N=50}
\label{fig: N=50}
\end{figure}

The number of coding blocks has the greatest impact on error performance as the message is only considered successful if all the messages are decoded correctly. This shortening of the number of blocks leads to a noticeable decrease in error rate performance. The overhead of half a block that exists for SWSC-based methods did not lead to as much of a performance change as one might expect. This matches with the assumptions made in the original SWSC paper \cite{SWSCOG} as it did not find a need to take into account the additional half-block length of code. 
The very small number of blocks shown in Fig. \ref{fig: N=20} shows a slight improvement over a medium number of blocks shown in Fig. \ref{fig: N=50} which performs slightly better than those in the benchmark with a large number of devices in Figure \ref{fig: OG}.

\subsection{The Impact of Block Ratio $\alpha$}
\begin{figure}[H]
\centerline{\includegraphics[scale=0.75]{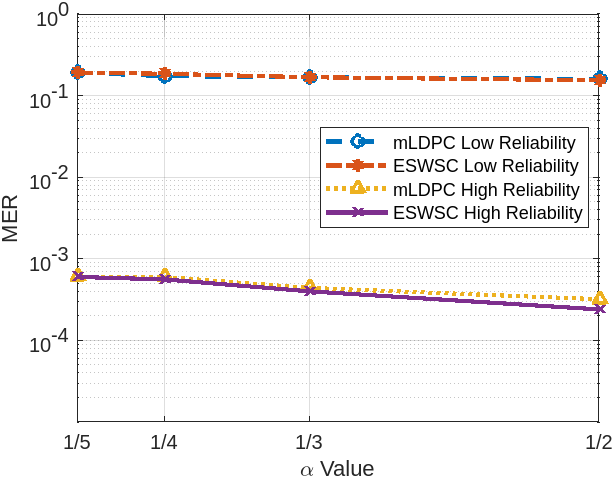}}
\caption{Values of \text{$\alpha$} at SNR=14 dB for low reliability, SNR=16dB for high reliability}
\label{fig: alpha graph}
\end{figure}

The $\alpha$ value is an important factor that can impact the results of the code. In order to ensure the requirements of URLLC are met for all information packets within the message, the $\alpha$ value is an adjustable value that can be modified for mMLC and ESWSC. The lowering of this value will lead to a decrease in error performance as observed in the following figure. The overall MER improvement from $\alpha=0.5$ to $\alpha=0.2$ is about 4-7\% in the regions of low-reliability requirements as shown in Fig. \ref{fig: alpha graph}. The drop is more evident in the high-reliability region on the log graph. It can be observed that in the low-reliability region, the mMLC outperforms ESWSC while the reverse is true for the high-reliability region. This suggests that ESWSC is able is a good choice for when there are high-reliability requirements like URLLC industrial automation. 

\section{Conclusion}
In conclusion, it can be observed that SWSC and ESWSC may not always yield the best results in all success rate regions and will usually struggle against their benchmark LDPC counterparts. In the regions of low MER rate that qualifies for URLLC, ESWSC, and SWSC are usually shown to have better results than their LDPC and mLDPC counterparts. When the number of blocks decrease, all the coding schemes tended towards better performance as expected. In cases of decreased block length, the SWSC-based coding schemes were able to maintain a slight performance improvement in the high-reliability requirement regions while losing out in the low-reliability requirement region. This trend continues into the comparisons with a varying $\alpha$ value as well where ESWSC outperforms mLPDC in the high-reliability region of BLER $\approx 10^{-6}$ but falls short in low-reliability requirement regions of BLER $\approx 10^{-3}$.

\bibliographystyle{IEEEtran}
\bibliography{sources}

\end{document}